\documentclass[aps,showpacs]{revtex4}% revtex
\usepackage{epsfig} %latex2e
\usepackage{graphics}%latex2e

\newcommand{\ket}[1]{| #1 \rangle} %revtex

\begin{document}

\title{A cesium gas strongly confined in one dimension : sideband cooling and
collisional properties}

\author{I. Bouchoule}
\author{ M. Morinaga}
\altaffiliation{Present address :
Institute for Laser Science
University of Electro-Communications
Chofu-city, Tokyo, 182-8585 Japan}
\affiliation{Laboratoire Kastler Brossel, 24 rue Lhomond, 75231,
Paris, France}
\author{D.S. Petrov}
\affiliation{
${}^1$ FOM Institute for Atomic and Molecular Physics, Kruislaan 407,
1098 SJ Amsterdam, The Netherlands \\
${}^2$ Russian Research Center, Kurchatov Institute, Kurchatov Square,
123182 Moscow, Russia
}
\author{C. Salomon}
\affiliation{Laboratoire Kastler Brossel, 24 rue Lhomond, 75231,
Paris, France}

\date{\today}


\begin{abstract}
 We study one-dimensional sideband cooling of Cesium atoms  strongly
confined in a  far-detuned optical lattice. The Lamb-Dicke regime is
achieved in the lattice direction 
%of oscillation frequency  
whereas the transverse confinement is much
weaker. The employed sideband cooling method, first studied  by
Vuletic et al.\cite{Vule98}, uses Raman transitions between Zeeman
levels and produces a spin-polarized sample. We present a detailed
study of this cooling method and investigate  the role of elastic collisions in
the system.  We accumulate $83(5)\%$ of the atoms in the vibrational
ground state of the strongly confined motion, and elastic collisions
cool the transverse motion to a temperature of
$2.8\,\mu $K=$0.7\,\hbar\omega_{\rm osc}/k_{\rm B}$, where $\omega_{\rm osc}$ is
the oscillation frequency in the strongly confined direction. 
The sample then approaches the regime of a quasi-2D cold
gas.  We analyze the limits of this cooling method and propose a
dynamical change of the trapping potential as a mean of cooling 
the atomic sample
to still lower temperatures. Measurements of the rate of thermalization 
between the weakly and strongly confined degrees of freedom are
compatible with the zero energy scattering  resonance observed
previously in weak 3D traps. For the explored temperature range
the measurements agree  
with recent calculations of quasi-2D collisions\cite{Petr01}. 
Transparent
analytical models reproduce the expected behavior for $k_{\rm
B}T \gg \hbar \omega_{\rm osc}$  and also for  $k_{\rm B}T \ll \hbar \omega_{\rm osc}$
where the  2D features are prominent.

\end{abstract}

\maketitle

\section{Introduction}

The physics of systems with reduced spatial dimensionality has  attracted
a great deal of interest.  These systems, where one (two)
degree(s) of freedom are confined to the quantum ground state of
motion, have  properties which can markedly differ from those of 3D
systems. The case of a 2D Bose gas has been explored theoretically
and experimentally with atomic hydrogen adsorbed on liquid helium
(see \cite{LesHouches92.Walraven,Safo98}). 
Advances on laser cooling  and Bose-Einstein
condensation of  atomic gases\cite{Varena98} have opened an interesting
possibility to create  new 1D and 2D quantum degenerate systems
\cite{Petr00,Petr00_2}. In a recent paper
Petrov et al. showed that  collisional properties of cold gases
can be drastically  modified by strongly confining the motion of
particles in one direction\cite{Petr00} : for instance, in a gas
with negative scattering length $a$, the mean-field interaction
can switch sign under variations of  the strong
confinement. This is of particular interest for the case of cesium.
Due to a very large and negative scattering length $a=-138\,nm$
 in the $\ket{F=3, m=3}$ state\cite{Leo00}, a cesium
condensate in a weakly confining trap would be unstable for a
few tens of atoms\cite{Rupr95}.
The characteristic size of the quantum ground state in
the strongly confined direction is  another important length in this
problem. For  a harmonic potential
with oscillation frequency $\omega_{\rm osc}$ this length is
$l_0=\sqrt{\hbar/2m \omega_{\rm osc}}$.
 Ref.\cite{Petr00} predicts that for a
sufficiently large value of the ratio $a/l_0$ the mean-field
interaction becomes positive allowing the formation of a stable
condensate.  A second interest in a strong confinement in 
one direction is
the possibility  to use a very efficient optical cooling method,
the sideband cooling \cite{Died89,Perr98,Bouc99,Hamm98,Vule98,Han00}.
This method requires $l_0\leq \lambda$, (Lamb-Dicke regime) where
$\lambda$ is the optical wavelength of the  cooling transition. For
instance in ref.\cite{Bouc99}, $92(5)\%$ of the atoms have been
cooled to the ground state in the tightly confined direction.
Several trap configurations for achieving 2D gases are under
investigation. Many of them are based on  far-off resonance
optical dipole traps \cite{Grim00} and evanescent fields near the
surface of a dielectric \cite{Gauc98,Hamm00}. Another common
approach is to confine atoms in micro-wells of a 1D far-detuned
optical lattice \cite{Vule98,Bouc99,Sche00}.

 In this  article, we present  an implementation of sideband 
cooling of cesium atoms
in an intensity lattice created at the intersection of two far
detuned YAG laser beams. The cooling method, first studied in
ref.\cite{Vule98}, uses Raman transitions between Zeeman substates
of $\ket{F=3}$ and produces an atomic sample polarized in
$\ket{F=3,m=3}$. We give a detailed account of  cooling in our
 geometry which is different from that in ref.\cite{Vule98}. We
recall the cooling principle, analyze the  role of
the beam polarizations and discuss 
the steady-state of the one-dimensional
cooling. The intensity lattice is vertical but atoms are also weakly
confined in the horizontal plane owing to the YAG gaussian beam
intensity profiles.  Elastic collisions couple the vertical and
horizontal degrees of freedom so that  1D sideband cooling
efficiently cools in the three directions\cite{Vule98}. 
A horizontal temperature
$T_h=2.8\,\mu $K$=0.7\hbar\omega_{\rm osc}/k_B$ has been obtained,
which already corresponds to the condition of  a quasi
two-dimensional gas. 
We discuss the limits of
this cooling. In particular, we show that the cooling efficiency
becomes exponentially small for
 $k_{\rm B} T < \hbar\omega_{\rm osc}$ and propose a dynamical 
method to reach a lower ratio
$k_{\rm B} T / \hbar\omega_{\rm osc}$. Finally, we address the
question of the influence of the strong confinement on 
collisional properties.
%the modification of collisional properties due to the
%strong confinement. 
We measure the rates of thermalization between the
strongly and weakly confined degrees of freedom. In the explored
temperature range from $4$ to $20\,\mu $K, our results are compatible with
the zero energy scattering resonance previously observed in weak
3D traps \cite{Guer98.res33,Hopk00}. The experimental data 
also agree with calculations of
ref.\cite{Petr01} which take into account the quantum character
of the particle motion  in the strongly
confined direction. 
We propose simple
 analytical models that reproduce the expected behaviors 
for $k_{\rm B}T \gg \hbar
 \omega_{\rm osc}$,   and also for  $k_{\rm B}T \ll \hbar
 \omega_{\rm osc}$ where the
  2D character of the particle motion is prominent.

\section{The far detuned trap}

\begin{figure}[ht]
\centerline{\includegraphics{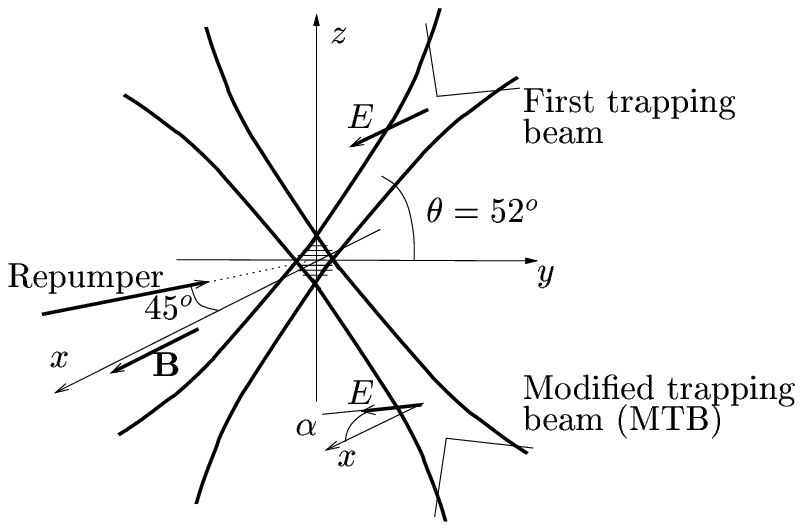}}
\centerline{\includegraphics{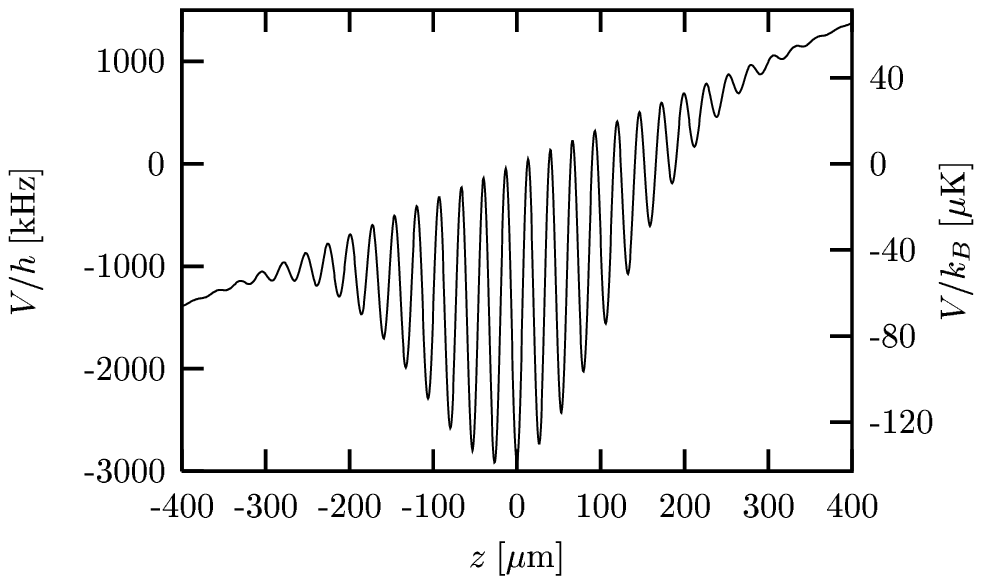}}
\caption{\it (a) : Laser beam configuration used in sideband cooling
of cesium.  Atoms are confined  in independent horizontal
microwells  resulting from the interference between the two YAG laser beams
propagating in the  $yz$ plane. A weak magnetic field is applied along
$x$. A repumping beam tuned to  
$|6s_{1/2},F=3\rangle\rightarrow \ket{6p_{3/2},F'=2}$ 
propagates in
the $xy$ plane and makes an angle of $45^\circ$ with the $x$ axis. The
polarization of one YAG beam is linear along $x$ and the polarization
of the other YAG beam (MB) makes a small angle $\alpha$ with $x$. (b) :
Trapping potential along $z$. For more visibility, the period of the
lattice (665\,nm) has been increased by a factor 40.}
\label{fig.trap}
\end{figure}

Our 1D intensity lattice is similar to the one
described in \cite{Bouc99}.  It is produced at the crossing of two
beams of a YAG laser (1.06\,$\mu $m), propagating in a vertical plane and
making an angle $\theta_{\rm YAG}=52^{\rm o}$ with the horizontal
direction. Both beams have linear  polarisations as presented in
fig. \ref{fig.trap}.  Each beam has a waist of 100\, $\mu $m and a power
of about 5\, W.  The YAG laser is far detuned to the red 
%($\Delta\gg \Gamma$) 
of the D1 and D2  transitions of cesium at 852\,nm and
894\,nm.  The atoms experience  a dipole potential proportional to
the laser intensity that confines the atoms in 
regions of maximum intensity.  For horizontal polarizations, the 
interference of
the two beams creates a vertical intensity lattice with a period of  665\, nm,
and the dipole potential is modulated in the vertical direction.  
The horizontal confinement is provided by the gaussian
shape of the YAG beams.  The total trap depth is about
140\,$\mu $K. Tunneling between different micro-wells is totally
negligible for temperatures smaller than $50\,\mu $K and the atoms are
confined in independent micro-traps.  In the central micro-trap, the
vertical oscillation frequency is  $\omega_{\rm osc}/2\pi=80\,$kHz 
and the horizontal
oscillation frequencies are $\omega_x/2\pi=175\,$Hz and
$\omega_y/2\pi=140\,$Hz.  The spontaneous emission rate at the bottom
of the trap is about 3\,s$^{-1}$. It leads to a heating rate of
$0.38\,\mu $K/s that we can neglect in the following experiments.  This
trap is loaded from a magneto-optical trap as explained in
\cite{Bouc99}.  About $2\times 10^5$ atoms are trapped  in a
gaussian cloud of rms dimensions $\sigma_x=31\,\mu $m, $\sigma_y=39\mu $m,
$\sigma_z=60\mu $m, thus populating about 200 horizontal micro-traps. Their
temperature is $\sim 20\,\mu $K, which for the vertical motion
 corresponds to a mean vibrational
number $\langle n \rangle =5.8$. The vertical oscillation frequency
depends on the position in the trap and
varies by 15\% within the spatial extension of the cloud. 
%the spatial extension of the cloud leads to a spread of 15\% of 
%the distribution of vertical oscillation frequencies.
%varies by 15\% on a distance scale of the spatial extension of the
%cloud.  

Using a charged-coupled-device camera, we perform two-dimensional
absorption images of the atomic cloud. The probe beam is horizontal
and makes an angle of $45^{\rm o}$ with the plane of the YAG beams.
First, an image taken just after  switching off the YAG beams gives a
measurement of the trap size. Second, an image taken after a free
expansion time of the cloud gives access to the velocity distribution
in the vertical and  horizontal directions in the plane of
the camera. This time of flight method is 
detailed in \cite{Mori99}.  

The
Lamb-Dicke parameter is  $\eta=\sqrt{\omega_{\rm rec}/\omega_{\rm osc}}=0.16$
where $\omega_{\rm rec}=\hbar k^2/2m$ is the recoil frequency
associated with the D2 transition.
 This low value enables us to cool the atoms to the ground
state of motion in the vertical direction by using sideband cooling
methods \cite{Perr98,Hamm98,Bouc99,Vule99}.

\section{ Cooling method}

\subsection{Principle}

  We describe here the main elements of the sideband cooling of the
vertical motion as developed by   Vuletic et al.\cite{Vule98}. 
This cooling uses only Zeeman sublevels of the Cs $F=3$ hyperfine 
ground state and produces a 
sample polarized in $\ket{F=3,m=3}$.
A
magnetic field is applied along the $x$ axis 
such that the Zeeman sublevels of  $F=3$ 
acquire different energy shifts.
  A repumping
laser which is resonant with the transition $\ket{6s_{1/2}, F=3}\rightarrow
\ket{6p_{3/2},F'=2}$ and which has a polarization with only  $\pi$ and
$\sigma_+$ components, provides  a finite linewidth of all Zeeman
substates of $F=3$ except for $\ket{F=3,m=3}$. Zeeman states with $m$
differing by $\pm 1$ are coupled by a Raman process such that the Raman
transition $\ket{m}\rightarrow \ket{m-1}$ is resonant with the
transition that decreases the vibrational quantum number by one.  
If the Raman
coupling and the intensity of the  repumping laser are sufficiently
small, the motional sidebands $\ket{m=3,n}\rightarrow\ket{m=2,n'}$ are
well resolved and the state $\ket{m=3,n=0}$ is a quasi-dark state. Its
lifetime is limited only by off-resonant Raman transfer and is much longer
than that of the other states.  In the Lamb-Dicke regime the
vibrational energy of the atoms is much larger than the recoil
energy. Therefore, the Raman transition to
a Zeeman level with $m<3$ followed by optical repumping  to $m=3$,
 leads on average to a decrease of the atoms' energy.
 Thus, atoms are cooled by repetition of such
cycles and they accumulate in the quasi-dark state $\ket{m=3,n=0}$.  A weak
intensity laser tuned to the  $\ket{F=4}\rightarrow\ket{F'=4}$ transition
prevents  atoms from accumulating  in the $\ket{F=4}$ state after spontaneous
emission.
The magnetic field is chosen so that the energy of $\ket{m,n}$ equals
the energy of $\ket{m-1,n-1}$. The Raman transition is then resonant
if the
two Raman beams have the same frequency, and
in contrast to our previous work\cite{Bouc99,Perr98}
 we simply use the YAG
beams themselves to induce the Raman transitions.  
In order to introduce  coupling
between different Zeeman levels, the polarization of one of the beams
is slightly modified from the linear horizontal polarization (see
section \ref{YAG}).
%As the detuning of the YAG laser with respect to the 
%D1 and D2 transitions 
%is much larger than the natural linewidth of the $6P_{3/2}$ and 
%$6P_{1/2}$ states, spontaneous emission 

\subsection{Repumping beam} 

The repumping beam tuned to  the 
$\ket{F=3}\rightarrow \ket{F'=2}$ transition of the D2 line propagates 
in the $xy$ plane and makes an angle of
$45^\circ$ with the $x$ axis. 
With respect to the axis of the magnetic
field , the total intensity of the beam $I$  splits in $I/3$ for 
the $\pi$
polarization (i.e. along the axis of the magnetic field), 
and $2I/3$ for the $\sigma^+$ polarization.  
%The polarization is described with respect to the magnetic field 
%axis as 

The population
of the atoms in each of the  Zeeman sublevels of $F=3$ is measured
via a frequency selective Raman transitions to $\ket{F=4}$. 
For this purpose we use
two Raman beams with frequencies differing by about 9.2 GHz 
as explained
in \cite{Perr99}.  

\begin{figure}
\centerline{\includegraphics{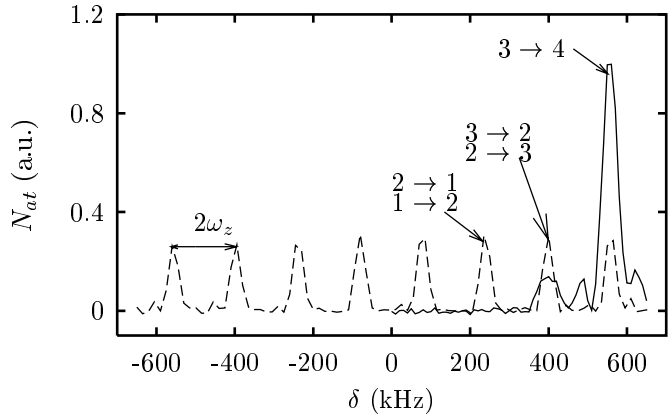}}
\caption{ Number of atoms transferred to the hyperfine ground state
$\ket{F=4}$ by a two-photon Raman pulse as a function of the detuning
of the two-photon transition from 9.192 631 770\,GHz.
The Raman beams propagate in the same direction 
in order to avoid change of vibrational state during the transfer.
 The labels
indicate the various $\ket{F=3,m}\rightarrow\ket{F=4,m'}$ transitions.
Dashed line~:  unpolarized atoms.  Solid line~: atoms polarized by a
3.6\,ms repumping laser pulse with an intensity of 0.3\,$I_{\rm
sat}$. The resulting polarisation exceeds 90\%. }
\label{fig.mesB}
\end{figure}

 With this method we  optimize the parameters of the repumping beam
in the absence of Raman coupling between Zeeman
sublevels i.e. with parallel linear polarizations of 
the YAG beams (see section \ref{YAG}).
 We can measure both the polarisation time constant and  the
steady-state polarization. 
With our experimental precision, the equilibrium population in
$\ket{m=3}$ deduced from the data in fig. \ref{fig.mesB}
is between $90\%$  and $100\%$. 
 We then deduce an intensity for the $\sigma_-$ component of the repumping
laser smaller than 3\% of the $\pi$ intensity.  The measured $1/e$
polarization time of an initially unpolarized sample is
$10\,\mu $s$\times I_{sat}/I$ at small intensities. The calculated lifetime
$\Gamma'^{-1}$ of the state $\ket{m=2}$ is $0.8\,\mu $s$ I_{sat}/I$.
The optical density of our sample is about 2 and we have seen no
influence of the number of atoms on the polarization time and on the
equilibrium state. This means that we see no effect of reabsorption
of photons on the polarisation.

\subsection{Polarization of the YAG beams}
\label{YAG}

\begin{figure}
\centerline{\includegraphics{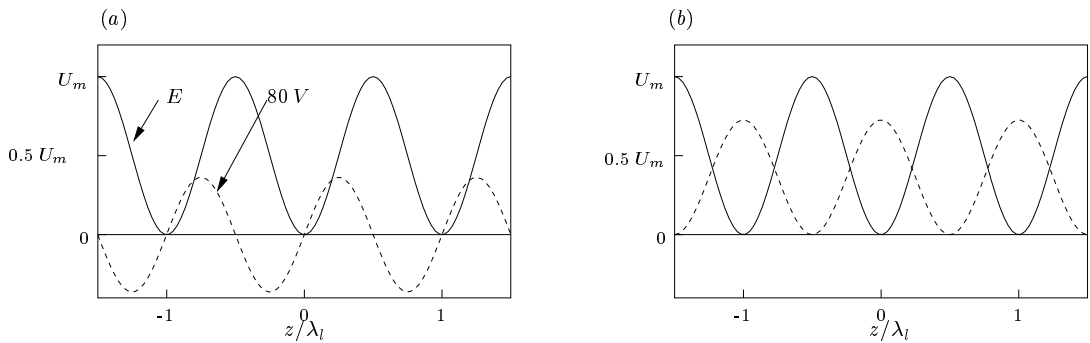}}
%\centerline{\input{couplageYAGpolar-art.pstex_t}}
\vspace{0.5 cm}
\caption{\it Raman coupling $V(z)$ ( dashed line) between
$\ket{F=3,m=3}$ and $\ket{F=3,m=2}$ as a function of the vertical
position of the atom. Figure ($a$) corresponds to a linear
polarization of the MB making an angle of $20^{\rm o}$ with the
horizontal direction and figure ($b$) corresponds to a polarization of
the MB with the same proportion of horizontal polarization but with a
relative phase of $\pi/2$ between the two polarizations.  The solid
lines depicts the dipole potential  seen by the atoms, the potential
being chosen 0 at the bottom of the micro-wells. For better visibility,
we have plotted $80\times V(z)$.  }
\label{couplageYAG.fig}
\end{figure}

The YAG
detuning is much larger than the hyperfine splittings
of the $6s_{1/2}$ and $6p_{3/2}$ states.
 Hence, if 
the two YAG beams are linearly polarized along $x$, there is no
coupling between different Zeeman sub-levels. The light-shift operator
is scalar (proportional to the identity operator) and all Zeeman
substates see the same YAG potential \cite{Grim00}.  Raman coupling
between different Zeeman levels is obtained by introducing, for one of
the trapping beams (called MB), a small component of polarization $X^0$
orthogonal to the horizontal one. All Zeeman sublevels  still share a
common potential energy and levels $|m,n\rangle$ and $|m-1,
n-1\rangle$ are degenerate for any $m\in[-2,3]$ and vibrational level 
$n> 0$. Raman transfers enable
a change of the vertical motion. This corresponds to absorption from one
beam and stimulated emission into the other beam. Due to an interference
between the two different processes, at the bottom of a micro-trap
the Raman coupling depends on the relative phase between the component
of horizontal polarization and the component along $X^0$.  The
coupling is even in $z$ for a phase difference of $\pi/2$ ( most elliptical
polarisation of the MB) but odd for a zero phase difference (linear
polarisation of the MB) as depicted in fig. \ref{couplageYAG.fig}.
Thus, due to the parity properties of the vibrational states,  linear
polarization of the MB induces the maximum coupling between neighboring
vibrational level whereas a phase difference of $\pi/2$ between the
polarization components induces a coupling only between vibrational
levels of the same parity.  In the experiment, the relative phase
between the two polarization components of the MB is controlled by an
adjustable retardation wave plate.  For a linear polarization of the
MB making an angle $\alpha$ with respect to  $0x$ (see fig.
\ref{fig.trap}), the coupling between
$\ket{m=3,n}\rightarrow\ket{m=2,n-1}$ is
\begin{equation}
V=\frac{\sqrt{6}}{24}\eta U_0\Delta_{\rm YAG}\left (
\frac{1}{\Delta_1}
-\frac{1}{\Delta_2}\right ) \sin (\alpha)  \sin(\theta_{\rm
YAG})\sqrt{n},
=V_R\sqrt{n}
\end{equation} 
where $\Delta_1$ (resp. $\Delta_2$) is the detuning of
the YAG beams with respect to the Cs $D_1 $ (resp. $D_2$) line, $\Delta_{\rm
YAG}=\Delta_1/3 +2\Delta_2/3$ and $U_0=4 (E_0d_0)^2/\Delta_{\rm YAG}$
is the depth of the trap with linear polarizations. In this formula,
only the component of order 1 in $\eta$ has been retained, which is a
good approximation in our situation where $\eta=0.16$. For an angle
$\alpha = 20^{\rm o}$, the Rabi frequency of the coupling
$\ket{m=3,n=1}\rightarrow\ket{m=2,n=0}$ is
\begin{equation}
\Omega_R=2\frac{V_R}{\hbar}=2\pi\times 6\,{\rm kHz}.
\label{eq.couplage_YAG}
\end{equation}

\begin{figure}
\includegraphics{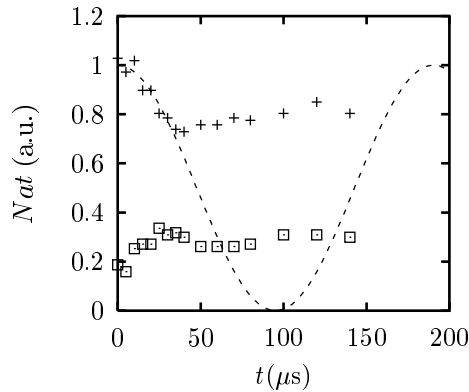}
\caption{Experimental determination of the YAG coupling 
$\ket{F=3,m=3,n}\rightarrow \ket{F=3,m=2,n-1}$.
The atoms are previously polarized in $\ket{F=3,m=3}$ by a strong
pulse  of the repumping beam.
Crosses : evolution of the number of atoms transfered from
$\ket{F=3.m=3}$ to $\ket{F=4,m=4}$ as a function of the delay time 
of the   Raman probe pulse.
Dashed line : fit of the first 40\,$\mu $s with a sine function of
amplitude 1.
Squares : number of atoms transfered to $\ket{F=4}$ on the transitions
$\ket{F=3,m=3}\rightarrow \ket{F=4,m=2}$ and 
$\ket{F=3,m=2}\rightarrow \ket{F=4,m=3}$ (see fig. \ref{fig.mesB}
and ref. \cite{Perr99}).  
We deduce a Rabi frequency $\Omega_{R}\simeq 5$\,kHz.
}
\label{fig.depolar}
\end{figure}

 In order to measure the coupling due to the YAG beams, we
first polarize the atoms in $\ket{F=3,m=3}$ by an intense repumping
pulse.
We then measure the time evolution of the $\ket{F=3,m=3}$ population
using a Raman transition to $\ket{F=4,m=4}$ as in fig. \ref{fig.mesB}
(see also ref. \cite{Perr99}).
 Figure \ref{fig.depolar} gives the
evolution  of the population  in 
$\ket{F=3,m=3}$.
The magnetic field for this experiment is chosen so that the levels
$\ket{m=3,n}$ and $\ket{m=2,n-1}$ are degenerate and the angle of
polarization of MB is $\alpha=20^{\rm o}$.
 We observe a damped oscillation.  A damping with a time constant 
 $\tau\simeq\sqrt{\hbar\omega_{\rm osc}/(k_{\rm B}T)}\Omega_R$ is expected 
as the coupling 
$\ket{m=3,n}\rightarrow \ket{m=2,n-1}$ is proportional to $\sqrt{n}$.
For a temperature of  $26\,\mu $K, we expect 
$\tau\simeq 0.4 \Omega_R$. 
 The Raman coupling, averaged over the population of the vibrational
states, is obtained by fitting the first 40\,$\mu $s of the oscillation
with a sine function of amplitude 1.
We obtain a Rabi frequency of 5\,kHz
in reasonable agrement with eq. \ref{eq.couplage_YAG}.
 In this analysis we have supposed that all  atoms 
contribute to the oscillations. However, this is not 
strictly the case.
First atoms in $\ket{m=3,n=0}$ are not transfered to $\ket{m=2}$. 
Second,  atoms
far from the center of the trap have a vertical
oscillation frequency which differs from the central one.
When the difference exceeds $\sim$5\,kHz, the coupling is no longer
perfectly resonant. 
 Both of  these
effects would lead to a  Rabi frequency slightly higher than
the estimation of 5\,kHz.   

\section{Cooling of the vertical motion}

\subsection{Measured temperature}

\begin{figure}
\centerline{\includegraphics{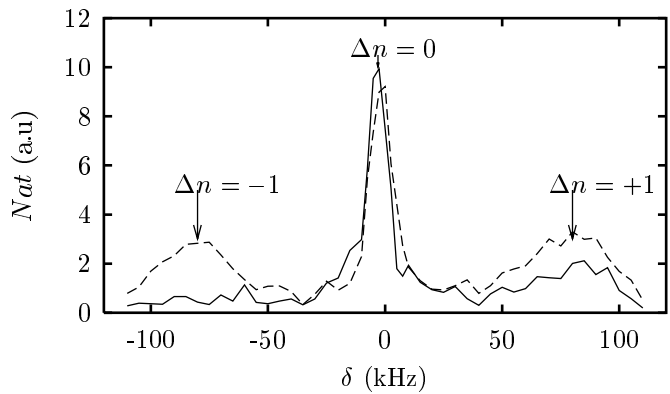}}
\caption{
Number of atoms transfered to the hyperfine ground state $\ket{F=4}$
by a two-photon Raman pulse as a function of the detuning of the 
two photon transition from 9.192631770 GHz. The two Raman beams are
counterpropagating in order to enable change of vibrational level.
The magnetic fiel has been turned off before the Raman pulse so that 
all the transitions $\ket{F=3,m}\rightarrow\ket{F=4,m'}$ are
degenerate.
Solid line : spectrum taken after 20\,ms cooling. 
Dashed line : spectrum taken before cooling.
The labels indicate the change of vibrational level during the transfer.
 On the spectrum taken after cooling, contrary to the 
spectrum taken before cooling, the
sideband corresponding to the transfer with a decrease of the
vibrational level by 1 has deasapeared. This is expected if all the atoms 
 are in the motional ground state $\ket{n=0}$.
}
\label{sideband}
\end{figure}

The smallest vertical temperature we achieved with this sideband
cooling method is about $0.56\hbar\omega_{\rm osc}$, which corresponds
to about 83(5)\% of the atoms accumulated in the ground state of
motion. This temperature is measured either by time of flight
absorption imaging or by using a Raman transition from
$\ket{F=3,m=3}$ to $\ket{F=4,m=4}$ with counterpropagating beams. The
vibrational sidebands are resolved (fig.\ref{sideband}) and from the
height of the red sideband we deduce the population of the $n>0$
vibrational states \cite{Bouc99}. Both measurements are in agreement
at the $5\%$ level.

The measured $1/e$ cooling time constant is 2.5\,ms. 
Such a cooling is obtained
for an intensity of the repumping beam of $2.4\times  10^{-2}I_{\rm sat}$, 
where $I_{\rm sat}=1.1\,$mW/cm$^2$ is the saturation intensity.
This beam intensity 
corresponds to a calculated energy width of $\ket{m=2}$
$\hbar\Gamma'=h\times 4.8\,$kHz and a linear polarization of the MB with
$\alpha=20^{\rm o}$ corresponding to a Raman coupling
$\ket{m=3,n=1}\rightarrow\ket{m=2,n=0}$ of $h\times  2.5\,$kHz. This ground
state population of $83\%$ is slightly less than the value ($92(5)\%$)
we previously obtained with sideband Raman cooling of unpolarized
atoms involving the two hyperfine states $\ket{F=3}$ and $\ket{F=4}$
\cite{Bouc99}.

\subsection{Discussion}

 The expected  equilibrium temperature can be easily calculated in the
Lamb-Dicke regime where $\eta \ll 1$ if the equilibrium is set by
non-resonant excitation from $\ket{m=3,n=0}$ to $\ket{m=2}$,
\cite{Wine79,Java84,Lind84}. The following simple argument gives the
equilibrium temperature in the limit where the Raman coupling is much
smaller than the lifetime $\Gamma'$ of $\ket{m=2}$.  For $\eta \ll 1$, we
can neglect any change of  vibrational level during the repumping
process. Furthermore, the Raman transfer inducing a change of the
vibrational level by $k\geq 2$, of order $\eta^k$, can be neglected
compared to the Raman coupling with a change of vibrational level by
1. Thus, the two processes in competition that define the equilibrium
are the process
$\ket{m=3,n=0}\rightarrow\ket{m=2,n=1}\Rightarrow\ket{m=3,n=1}$ and
the resonant process $\ket{m=3,n=1}\rightarrow
\ket{m=2,n=0}\Rightarrow\ket{m=3,n=0}$, where $\Rightarrow$ indicates
optical pumping. Note that the Raman transfer
$\ket{m=3,n=0}\rightarrow \ket{m=2,n=0}$ is not allowed with linear
polarization of the MB.  In the limit where the Raman coupling is much
smaller than the energy width $\Gamma'$ of $\ket{m=2}$, equaling 
the rates of these two processes gives a relative population of
$\ket{m=3,n=1}$ of $(\Gamma'/4\omega_{\rm osc})^2$.  The expected
cooling rate is $\Gamma_{\rm cool}=\Omega_R^2/\Gamma'$.  With our
experimental parameters, this  leads to a temperature of about
$0.12\hbar\omega_{\rm osc}$ (population of $\ket{m=3,n=1}$ of 0.0002),
a factor of 4.6 smaller than the obtained temperature. The predicted
cooling rate is $3\times 10^4$\,s$^{-1}$ ($1/e$ cooling time of 30\,$\mu $s).

We discuss now some possible sources of discrepancy between this
simple model and the experiment, although we find that they are not
sufficient to explain the observed equilibrium 
temperature.  
First, the resonance of the transition
$\ket{m=3,n=1}\rightarrow \ket{m=2,n=0}$ is not fulfilled for all the
atoms.  Indeed, the vertical oscillation frequency of the atoms, which
depends on their position in the trap, spreads over about 15\% of the
central 80\,kHz frequency.  If the transition is detuned by $\delta >
\Gamma'$, the expected steady-state population of $\ket{m=3,n=1}$ is
$(\delta/(2\omega_{\rm osc}))^2$. For $\delta=0.15\omega_{\rm osc}$,
this population is only 0.005, still smaller than the measured one.
However, the cooling rate is significantly modified by this effect;
for instance, it is reduced by a factor 36 for a detuning of
0.15$\hbar\omega_{\rm osc}=12\,$kHz.  
Thus, the spread of oscillation frequencies is likely to be 
the cause of the long cooling time observed.
Second, non resonant  Raman
transfer is not the only process causing a departure from
$\ket{m=3,n=0}$.  Indeed, a component of $\sigma_-$ polarization of
the repumping beam would excite atoms in $\ket{m=3}$. The measurement
of the steady-state polarization without Raman coupling shows that the
excitation rate of $\ket{m=3}$ is smaller than $0.01\Gamma'$.  The
heating produced by this excitation is smaller than or on the order of
$0.01 \times E_{\rm rec}\Gamma'$.  For atoms resonant with the
$\ket{m=3,n=1}\rightarrow \ket{m=2,n=0}$, this would lead to a steady
state population in the excited vibrational states of 0.003 much
smaller than the measured one. However, atoms detuned by 12\,kHz (the 
typical width of the distribution of oscillation frequency in the cloud) 
experience a smaller cooling rate and therefore the steady-state population
of excited states increases to  about 1\% for these atoms.

Finally, heating due to
multiple photon scattering within the cloud is likely to bring a
negligible contribution to the observed mean vibrational number
because the optical density is only 2.
%.
%Indeed, the recoil energy is much smaller than the oscillation
%frequency and the optical depth on resonance is only 2.

\subsection{Effect of the polarization of the MB}

\begin{figure}[ht]
%\centerline{\input{le200300_ib.tex}}
\centerline{\includegraphics{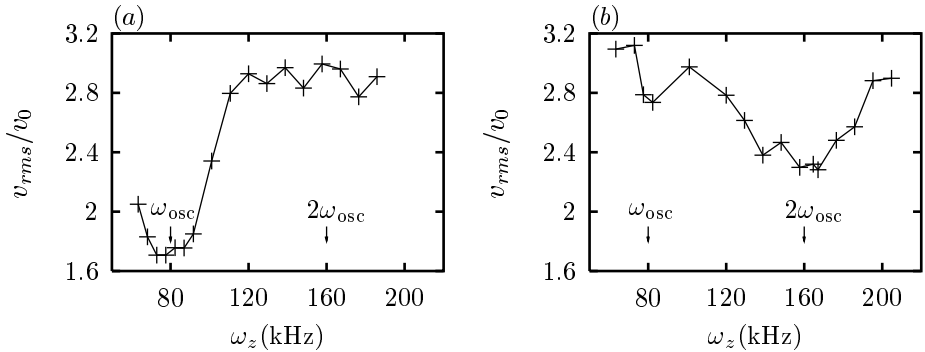}}
\vspace{0.2cm}
%\centerline{\input{resbarticle.tex}}%dans these/le200300
\caption{\it rms width of the vertical velocity distribution obtained
after 30\,ms of cooling as a function of the Zeeman splitting between
adjacent Zeeman levels. $v_0=\sqrt{\hbar m \omega_{\rm osc}/2}$ 
is the rms width
of the vibrationnal ground state. The polarization of the MB is linear
in ($a$) and elliptical in ($b$). }
\label{fig.lrv2n}
\end{figure}

 The  Raman transfer $\ket{m=3,n}\rightarrow \ket{m=2,n-1}$ is
resonant when the Zeeman splitting is equal to the vibrational
energy. Figure \ref{fig.lrv2n} ($a$) shows this resonant behavior
of the cooling as a function of magnetic field. For a Zeeman splitting
equal to $2\omega_{\rm osc}$, a transfer $\ket{m=3,n}\rightarrow
\ket{m=2,n-2}$ is resonant. But with a linear polarization of the MB,
such a transfer is forbidden (no coupling between levels of same
parity as shown in  fig. \ref{couplageYAG.fig} of section \ref{YAG}).
In contrast, when we choose an elliptical polarization of the MB,
cooling on the second sideband is observed (fig. \ref{fig.lrv2n}
($b$)). This cooling is weak because the Raman coupling is very small
for a small Lamb Dicke parameter. In this case, as expected, almost no
cooling is observed on the first sideband.

\section{Cooling the horizontal motion}
\label{sec.collhor}
\begin{figure}[ht]
%\centerline{\input{le060799_ref3Dlinparlin_2axes.tex}}
\centerline{\includegraphics{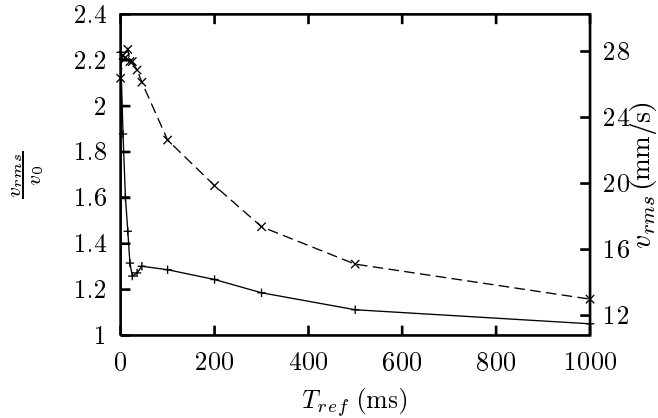}}
\vspace{0.5cm}
\caption{\it Evolution of the rms velocity width in the vertical (+)
and the horizontal ($\times $) directions as a function of the
cooling time.  $v_0$ is the width of the ground state of motion in the vertical direction.  Note, at very short cooling time the initial increase of the velocity spread in the horizontal direction. It is due to the spontaneously emitted photons required for
the sideband cooling of the vertical motion.
}
\vspace{0.5cm}
\label{fig.cool3d}
\end{figure}

To cool the horizontal degrees of freedom with the vertical sideband
cooling, a transfer of  energy from  horizontal to vertical
motion is needed.  This transfer can be provided by collisions between
atoms and efficient cooling of the three degrees of freedom has been
demonstrated in \cite{Vule98}.  In our experiments, after 1\,s of
vertical sideband cooling, the temperature drops from $T_h=T_v=13\,\mu $K
to $T_h=2.7\,\mu $K and $T_v=1.6\,\mu $K as presented in fig.
\ref{fig.cool3d}. For these data, the cooling rate of the vertical
motion is much higher than the horizontal one.  
%Such an energy
%transfer exists in our system as shown by the observation of a
%thermalization between the vertical and the horizontal motion
%following the application of the vertical cooling.  This transfer is
%provided by collisions between atoms and this can be used to cool the
%three degree of freedom. Indeed, we have cheked taht  the measured
%thermalisation rate between the horizontal and the vertical motion is
%about proportionnal to the number of trapped atoms.  With this
%coupling between the horizontal and the vertical motion, an
%application of the vertical sideband cooling on a time longer than the
%collisional time, can lead to a cooling of the horizontal motion.
%When we apply the cooling for a long time, we observe indeed a
%decrease of the horizontal temperature presented figure
%\ref{fig.cool3d}.  For this cooling, 
The intensity of  the repumping
beam is decreased to about $2.5\times 10^{-3}\,I_{\rm sat}$, which is about
10 times smaller than the power that leads to the most  effective
cooling of the vertical motion.  This is due to the heating of the
horizontal motion experienced by the atoms even after reaching the
steady state of the vertical motion.  Indeed, the steady state of the
vertical cooling is an equilibrium between heating processes like
excitation by the $\sigma_-$ component of the repumper  and the
cooling itself.  Spontaneous photons are emitted at a rate
proportional to the intensity of the repumping laser and the
spontaneous photons emitted in the horizontal directions are
responsible for a heating of the horizontal motion.  To achieve a
heating smaller than the cooling rate due to the collisions, the power
of the repumper should thus be small enough.  Figure \ref{fig.cool3d}
also indicates that the cooling in the vertical direction is affected
by the energy in the horizontal planes since the rms velocity in the
vertical direction tends slowly to the ground state rms velocity as
the horizontal temperature decreases.

 In the horizontal plane, we could  achieve a temperature $T=2.8\,\mu $K
($k_B T=0.7\hbar\omega_{\rm osc}$) which corresponds to a mean kinetic
energy per degree of freedom of $1.3\hbar \omega_{\rm osc}/4$. 
 For
such a low horizontal temperature, the energy transfer between the
horizontal and vertical degree of freedom due to the collisions is
very inefficient.  Indeed, we explain in section \ref{sec.inhib},
that for small horizontal temperatures, the rate of energy transfer
due to  collisions goes down exponentially  with a factor
$e^{-2\hbar\omega_{\rm osc}/(k_BT)}$ as the temperature decreases.
This factor, which is 0.5 for the typical initial horizontal
temperature ($k_{\rm B}T=3\hbar\omega_{\rm osc}$), is only 0.06 for the final
temperature.  Thus, even if the vertical equilibrium  temperature of
the sideband cooling $T_{\rm eq}$ is much smaller than
$\hbar\omega_{\rm osc}$, we expect that the rate of the horizontal
cooling  decreases exponentially as $k_BT<\hbar\omega_{\rm
osc}$. Small external sources of heating would then cause the
horizontal cooling to stop before $T_{\rm eq}$ has been reached.  This
inhibition of the coupling between the vertical and horizontal degrees
of freedom for temperatures smaller than $\hbar\omega_{\rm osc}$ is an
important limitation of  3D cooling using 1D sideband cooling and
collisions.

 The highest phase space density we reached is 
$n\lambda_{\rm DB}^3=1.3\times 10^{-3}$, 
where $\lambda_{\rm DB}=\hbar\sqrt{2\pi/mk_{\rm B}T}$ is 
the de Broglie wave length and $n$ is the peak density.
This is obtained after cooling to a temperature 
$k_{B}T_h=k_{\rm B}T_v=4.3\,\mu $K with  
about 450 atoms per micro-traps 
at a peak density  $n=4\times 10^{12}$ atoms/cm$^3$.

\section{A method to cool further~: change of polarization
angle $\alpha$}

 To overcome the decrease of the cooling efficiency of the horizontal
motion as the temperature decreases, we changed the parameters of the
trap during the sideband cooling. More precisely, the ratio of
the horizontal to the vertical oscillation frequencies is increased.
Thus,  if the change is done adiabatically with respect to the
oscillation of the atoms, the ratio of the energy of the horizontal
motion to the energy of the vertical motion will increase. More pairs
of colliding atoms will then have an horizontal energy sufficient to
populate the excited vibrational states after the collision.  This
change is realized by increasing the angle $\alpha$ of the
polarization of the MB with respect to the horizontal direction, using
a retardable wave plate on a time scale of about 10\,ms.
With a large angle $\alpha$, the contrast of the interferences between
the two trapping beams is reduced leading to a decrease of the
vertical oscillation frequency  by a factor $\sqrt{\cos(\alpha)}$.
Because the depth of the trap is also reduced, the horizontal
oscillation frequencies are also reduced but by a smaller factor of
$\sqrt{(1+\cos(\alpha))/2}$.

\begin{figure}[ht]
\centerline{\includegraphics{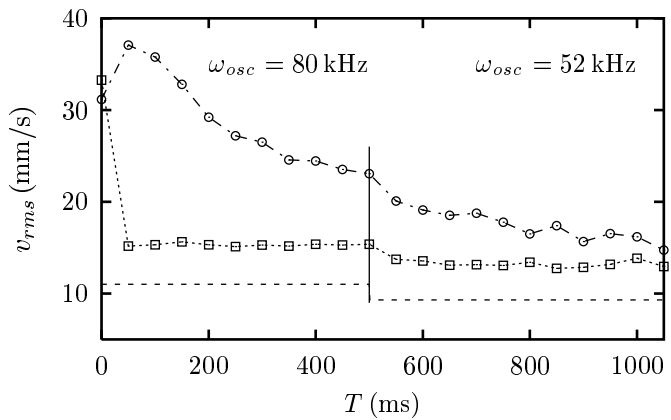}}
\vspace{0.5cm}
\caption{\it Two step cooling with a decrease of the vertical
oscillation frequency after 500\,ms.  squares~: rms velocity width
along z.  circles~: rms velocity width  along x.  The dashed
straight lines give the rms velocity  width of the ground state of
motion in the vertical direction.}
\label{fig.ref2etapes}
\end{figure}

 Experimentally, we decrease the  vertical oscillation frequency  from
$\omega_{z_1}=80(3)$ to $\omega_{z_2}=53(2)$\,kHz, (decrease by a
factor 0.67), corresponding to  $\alpha$ going from  $29^{\rm o}$ to
$63^{\rm o}$.  The horizontal oscillation frequencies   decrease only
by a factor $\omega_{x_2}/\omega_{x_1}=0.88$.  At 53\,kHz, sideband
cooling  still works well, after the corresponding change of the
magnitude of the magnetic field.  We also found that a decrease of the
repumping power by a factor 3.3 was required. 
Figure \ref{fig.ref2etapes} gives the time 
evolution of the horizontal and vertical velocity distributions before
and after the change of the vertical oscillation frequency. More
cooling   after the change of the oscillation frequency is clearly
visible.

If we assume that the cooling for a large value of $\alpha$ produces
$k_B T_h=\beta \hbar \omega_z$, when returning  the polarization to
its original value this final horizontal temperature is given by $k_B
T_h/(\hbar\omega_{z_1})=\beta (\omega_{x_1}/\omega_{x_2})
(\omega_{z_2}/\omega_{z_1}) $.  This way, an horizontal temperature
that could not be obtained by cooling only with $\omega_{z_1}$ because
of the decoupling of the horizontal and vertical motion at low
temperature could be produced.

\section{Dependence of the temperature on the number of atoms and atom losses}
\begin{figure}

\centerline{\includegraphics{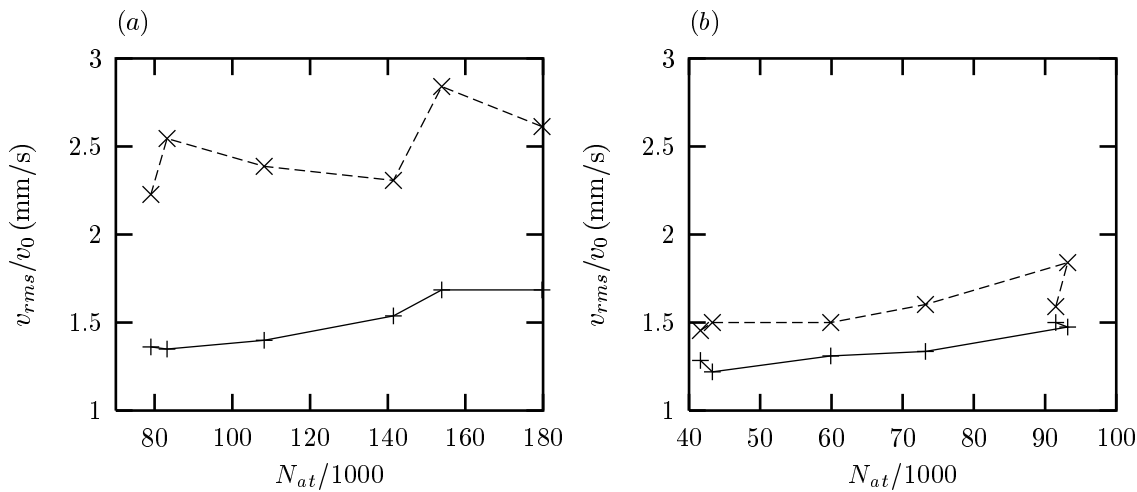}}
\caption{Width of the rms velocity in the horizontal (dashed line) and
vertical (solid line) direction after 60\,ms  ($a$) and 500\,ms ($b$)
cooling time as a function of the number of atoms measured at the 
end of the cooling.}
\label{fig.tempvsnat} 
\end{figure}

Dependence of the measured temperature with the number of atoms has
been observed as shown in fig. \ref{fig.tempvsnat}. 
At large atom numbers the temperature is higher.
This can be due to heating  by the reabsorption of 
spontaneous photons or by exothermic light-assisted collisions.

We also found that this sideband
cooling method is accompanied by losses of atoms~: about 50\% of the
atoms have been  lost in the first 500\,ms of the cooling  presented
in fig. \ref{fig.cool3d}.  The corresponding loss rate of about
1.0(1)\,s$^{-1}$ is significantly larger than the measured loss rate
due to collisions with the back-ground gas (0.4(1)\,s$^{-1}$).

 We believe that these losses are due to exothermic collisions
assisted by the repumper light\cite{Varena91.Juli}. 
However, a dependence of
the loss rate on the density of atoms characteristic of such a
2-body loss mechanism has not been demonstrated.  Indeed, due
to the dependence of the temperature on the number of atoms, only
small  changes  in the density have been achieved ($\sim 15\%$) and
no conclusive results could be deduced.

 For the depth of our trap (about 130\,$\mu $K), the dominant inelastic
light-assisted process is  radiative escape\cite{Galla89}. We
give below an estimate of the loss rate due to this process in our
experiments, based on earlier calculations made by Julienne and
Vigu{\'e}\cite{Juli91}. The loss rate can be written $Kn_{ex}$, where
$n_{ex}$ is the density of excited atoms.  For a temperature of
$100\,\mu $K, a trap depth of 1K, and a detuning of the laser of
$-\Gamma$, $K \simeq 0.2 \times 10^{-11}$\cite{Juli91}.  Because $K$ scales 
with the depth $U$ of the trap as
$U^{5/6}$ \cite{Juli91}, we expect $K$
to be  bigger by a factor $1.7\times 10^{3}$ in our trap of depth $130\,\mu $K.
In our experiment, the repumping beam is on resonance and we could
thus expect a change of the factor $K$.  However, we believe that it
will be of the same order of magnitude.  The density  of excited atoms
 after the first 10\,ms of strong vertical cooling,
is mainly determined by excitation of the atoms in $\ket{m=3}$ due to
the $\sigma_-$ component of the repumper.  With the experimental
parameters and taking a component of bad polarisation of the repumping
beam of 5\%,  $n_{ex}$ is about $n\times 0.6\times 10^{-4}$, where $n$ is the
total density of atoms.  With a mean density
$n=5\times 10^{11}$\,at/cm$^3$, the estimated
loss rate is then $0.15\,$s$^{-1}$.  This loss rate is of the same
order of magnitude as the measured one.  

We note that, contrary to
our findings,  no atom losses were observed in ref\cite{Vule98},
although the density achieved in \cite{Vule98} was 3 times higher than
ours.

\section{Collisional properties}

\subsection{Introduction}

 In the ground state of motion in the vertical direction,
atoms are confined in a gaussian distribution with   rms size
$l_0=20\,$nm.
 On the other hand,
the scattering length of cesium atoms which characterizes the
collisional properties at low energy is, for the $\ket{F=3,m=3}$
state, negative with an absolute value larger than 60\,nm
\cite{Hopk00,Vule99,Leo00} (The most recent calculations of
ref\cite{Leo00}, based on the analysis of several Feshbach
resonances of \cite{Vule99} give $a=-138\,$nm).
 An interesting question is then~: are the collisional
properties altered by the strong 1D confinement as compared to the
free space case ?
 With such a  high absolute value of the scattering length, for the
temperatures in our experiments, the collisional cross section for
free atoms reaches its maximum value of $8\pi/k^2$. This resonance
behavior has  been observed in weak traps by measuring
thermalization times as a function of temperature
\cite{Guer98.res33,Hopk00,Arnd97}.
 The   quantity $1/(nvT_{\rm therm})$, where
$n$ is the density of atoms,
 which is expected to be proportional
to the typical collision cross section $\sigma$, was found
proportional to $1/T$. This was interpreted as a zero energy
resonance and $\sigma$ was found equal to its maximum allowed
value of $8\pi/k^2$. We give below an analytical derivation, in
the classical limit ($k_B T\gg \hbar \omega_{\rm osc}$), of the
thermalization time $T_{\rm therm}$.

 In our trap, at  temperatures much larger than the vibrational
energy $\hbar\omega_{\rm osc}$, we expect the collisions to be 3D
collisions. Indeed, in this case, the motion of the atoms can be
understood as oscillations of wave packets of quite well-defined
wave vector $k$. The collision of two wave packets occurs on a
spatial scale $1/k$ (square root of the cross section).
For the typical $k=\sqrt{mk_B T}/\hbar$ and for $k_B T \gg 
\hbar\omega_{\rm osc}$, this spatial scale is much smaller than the
spatial amplitude of the oscillation $\hbar k /(m\omega_{\rm osc})$.

However, at  temperatures  $k_B T \leq \hbar \omega_{\rm osc}$, D. Petrov
and G. Schlyapnikov predict drastic changes of the collisional
and thermalization properties, such as a change of the sign of the
scattering length\cite{Petr00} and an exponentially vanishing rate
of the thermalization   between the strongly confined motion and
the transverse ones\cite{Petr01}.
 As a test of these possible changes we have investigated
the zero-energy resonance and searched for confinement-induced
modifications of the thermalization rates at various temperatures.

\subsection{Measurements}

A situation out of thermal equilibrium is easily produced by the
vertical sideband cooling method described above. As the cooling
rate of the vertical motion is much larger than the collision
rate, most of the atoms are quickly cooled to the ground state of
the vertical motion, while the horizontal motion is cooled more
slowly. At any cooling  time, the horizontal temperature is higher
than the vertical one.
 If the vertical cooling is stopped,
the cloud of atoms relaxes towards equilibrium. We then measure
the time  evolution of the widths of the velocity distributions
along the vertical direction and the horizontal direction. We
checked that this relaxation is indeed due to collisions by
measuring thermalization times for different numbers of atoms~:
$T_{\rm therm}$ is found to be inversely-proportional to the number of
trapped atoms. 

\begin{figure}[h!]
\centerline{\includegraphics{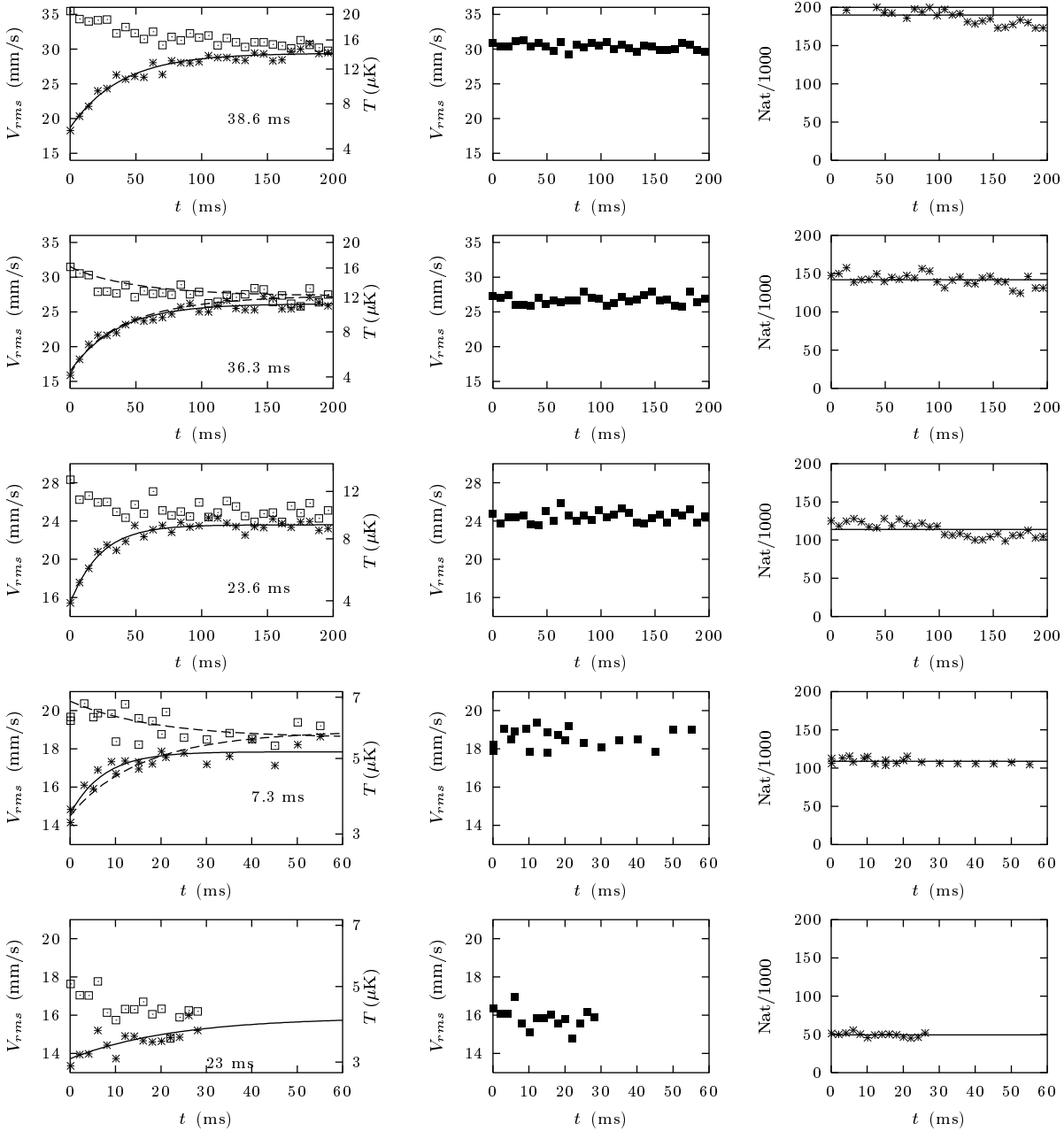}} \caption{\it Free evolution
of the rms width of the velocity distribution along the vertical
direction (stars,$v_z$) and along the horizontal direction $x$ (open
square, $v_{x}$), after switching off the sideband cooling for
different initial conditions. The solid line is an exponential fit of
the evolution of $v_z$ and the $1/e$ time is noted on the graphs.
On the second and fourth graph, the evolution predicted by the
calculations of D. Petrov and G. Shlyapnikov for the same initial
velocity distribution and for the same initial density are shown
in dashed lines \cite{Petr01}. The second column shows
$\sqrt{(v_z^2+2v_{x}^2)/3}$, the velocity corresponding to the
mean kinetic energy per degree of freedom. The third column gives
the number of atoms. Each point is the average of 20 successive
measurements.
}
\label{fig.exempletherms}
\end{figure}

The initial horizontal temperature is varied by
changing the time duration of the sideband cooling  phase before
the measure of thermalization. We could thus measure thermalization times for
initial horizontal temperatures varying from 20\,$\mu $K to
4\,$\mu $K. The maximum temperature  is limited by the depth  of
the trap.  Figure \ref{fig.exempletherms} presents 
many measurements of thermalizations corresponding 
to various initial temperatures. As
the absorption imaging is destructive, each point in fig.
\ref{fig.exempletherms} corresponds to a different experimental
cycle lasting about 2\,s. Fluctuations  in the number of atoms for
a given cooling time are   about 2\%.
We check that $2E_{c_x}+E_{c_z}$, where $E_{c_x}$ and $E_{c_z}$
are the mean kinetic energies along the vertical and horizontal
axis of the camera, is conserved to within about 5\% (see fig.
\ref{fig.exempletherms}). This behavior is expected if the
thermalization time is much larger than the oscillation time.
Indeed, in this case, we expect that the energy is always equally
distributed between the potential and kinetic energy and that the
horizontal velocity distribution stays isotropic.

\subsection{Results}

\begin{figure}[ht!]
%\centerline{\input{crossarticle.tex}}
 \centerline{\includegraphics{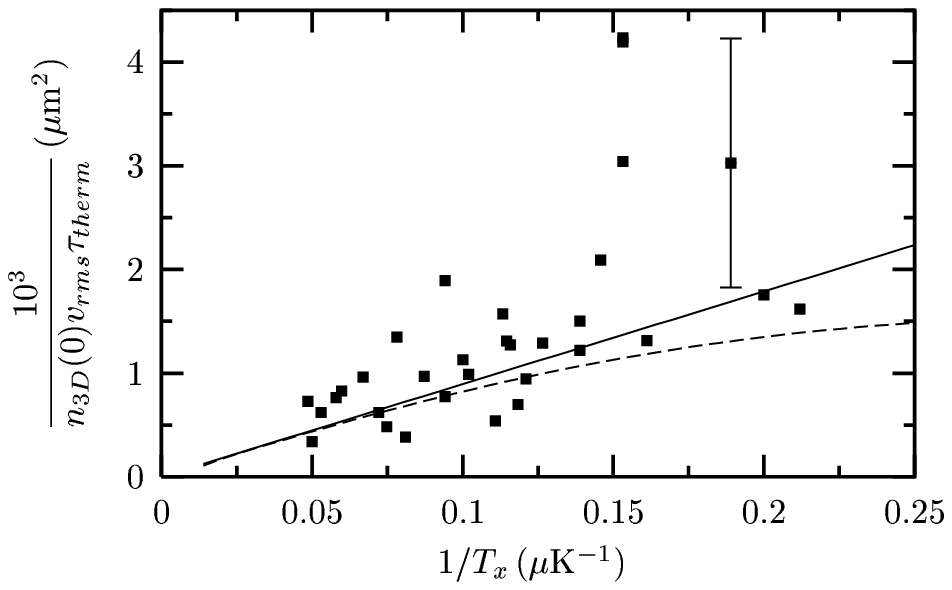}}
\caption{\it
Comparaison between experimental results of thermalization and
theoretical calculations in the temperature interval
$[4\,\mu $K$,20\,\mu $K$]$. The solid line corresponds to the 3-dimensional
behavior (eq. \ref{eq.3Dcoll}). The dashed line results from the
quantum calculations of D. Petrov et al. (see text). Two typical error
bars at high and low temperatures are shown.
.} \label{fig.collisionsExpetTheo}
\end{figure}

Figure \ref{fig.collisionsExpetTheo}, presents, as a function of
temperature, the quantity $1/(n_{3D}v_{rms}T_{\rm therm})$ where
$v_{rms}$ is the initial (i.e.  at the beginning of the
thermalization) rms width of the horizontal motion and $T_{\rm
therm}$ is the $1/e$ thermalization time deduced from an
exponential fit to the evolution of the vertical velocity width.
$n_{3D}$ is the initial mean density seen by an atom. It is
averaged over the different vertical micro-planes and the
population of the different micro-planes is deduced from the
measured gaussian vertical distribution of atoms and the lattice
period. Furthermore we assume that the temperature is identical in
all micro-planes. $n_{3D}$ is then deduced from the temperature
measurement and from the vertical and horizontal oscillation
frequencies. The horizontal oscillation frequencies (of the order
of $175$ and $140$\,Hz) are measured through the heating of the
atomic cloud produced by parametric
drive\cite{Landau.meca} with a precision of about 5\%.

 The straight line in fig.\ref{fig.collisionsExpetTheo} corresponds to
the expected behavior when $k_B T \gg \hbar \omega_{\rm osc}$. It results from
the 3D {\it classical} calculation presented below, where the
scattering resonance is assumed but where the quantization of the
vertical motion is not taken into account.  Within our experimental
accuracy and for the explored $4\,\mu $K-- $20\,\mu $K temperature range, we
have seen no modification of the scattering resonance induced by the
strong 1D confinement: our measurements are compatible with the {\it
classical} calculation.  They are also compatible with the full
quantum calculation performed by Petrov et al.\cite{Petr01}
(dashed line in fig.\ref{fig.collisionsExpetTheo}), which departs
significantly from the {\it classical} calculation only for
temperatures much below $4\,\mu $K. The quantum calculation assumes an
oscillation frequency of $80\,$kHz and a scattering length $a$ of
$-60\,$nm.  As this scattering length is already larger than the size
of the ground state wave function, the calculation does not differ
much from the $a=\infty$ case. It also shows that 
%deviation from
%the classical case only appears 
for $k_B T\ll \hbar\omega_{\rm osc}$
%where 
the thermalization time is expected to increase exponentially as
\begin{equation}
T_{\rm therm}=\frac{9m}{64\hbar}\frac{1}{\bar{n}_{2D}}
e^{\frac{\hbar\omega_{\rm osc}}{k_B T}}
\end{equation}

 As we will show below, this behavior can also be understood assuming that
the  scattering resonance is still valid but taking into account
the discrete structure of the vibrational energy in the vertical
direction and  parity conservation during the collision.

\subsection{Classical calculation of $T_{\rm therm}$ }

We propose here a calculation of the thermalization time in the
limit $k_B T \gg \hbar \omega_{\rm osc}$. In this condition, the motion of
the atoms can be treated classically. The collisions are the same
as for free atoms and can be considered as "point-like" because
the extension of the cloud is much bigger than the square root of
the typical cross-section. Finally we assume that
$\sigma=8\pi/k^2$.

 We are interested in the description of the thermalization between the
horizontal motion and the vertical one. Thus we consider the
evolution of
\begin{equation}
\Delta E = E_z-\frac{1}{2}E_{\rho},
\end{equation}
where $E_z$ and $E_{\rho}$ are the kinetic energies of the vertical and
horizontal motions. $\Delta E$ vanishes at thermal equilibrium. If we
neglect the anharmonicity of the trapping potential, only collisions
can change $\Delta E$. A collision where two atoms of velocities ${\bf
v_1}$ and ${\bf v_2}$  end up with the respective velocities ${\bf
v_3}$ and ${\bf v_4}$ changes $\Delta E$ by
\begin{equation}
\delta_{12\rightarrow 34}=\frac{1}{2}m \left (
v_{z_3}^2+v_{z_4}^2-v_{z_1}^2-v_{z_2}^2-\frac{1}{2}\left (
v_{\rho_3}^2+v_{\rho_4}^2-v_{\rho_1}^2-v_{\rho_2}^2 \right) \right ) ,
\end{equation}
where the indices $\rho$ and $z$ refer respectively to the horizontal
and vertical component of the velocity. Energy conservation implies
\begin{equation}
\delta_{12\rightarrow 34}=\frac{3}{2}\frac{1}{2}m \left (
v_{z_3}^2+v_{z_4}^2-v_{z_1}^2-v_{z_2}^2 \right ).
\end{equation}
If we note $f({\bf r},{\bf v})$ the phase space distribution, the rate
of change of $\Delta E$ is 
\begin{equation} 
\frac{d\Delta E}{dt}=\int
d^3{\bf r}\frac{1}{4} \int d^3{\bf v_1}d^3{\bf v_2}d^3{\bf v_3}d^3{\bf
v_4} \delta_{12\rightarrow 34} W_{12\rightarrow 34}\left ( f({\bf
r},{\bf v_1})f({\bf r},{\bf v_2})- f({\bf r},{\bf v_3})f({\bf r},{\bf
v_4}) \right ),
\label{eq.evol}
\end{equation}
where $W_{12\rightarrow 34}=W_{43\rightarrow 12}$ are the collision
rates.  The factor $1/4$ cancels the fact that each collision process
$(1\leftrightarrow 3,2\leftrightarrow 4)$ is counted 4 times.

Changing  variables in the integral, using the center of mass
velocity and the angle of deviation in the center of mass frame,
$W$ is
\begin{equation}
W_{12\rightarrow 34}=|{\bf v_2 -v_1}|\frac{\sigma}{4\pi}.
\end{equation}
with  $\sigma=8\pi/k^2$,
where $k=|{\bf v_2 -v_1}|m/(2\hbar)$ is the wave vector of the
relative motion.

 Equation \ref{eq.evol} alone does not give the evolution of the
distribution $f$. In general, one has to solve the Boltzmann
equation. But, in the collisionless regime where the time between
two collisions is much longer than the oscillation period of the
atoms in the trap,  the problem can be simplified. In this regime,
we do not expect oscillations of the cloud  and we can assume
that, for both the vertical and horizontal motion,
 the distribution $f$ corresponds to a Boltzmann
distribution   with temperatures $T_z$ and $T_{\rho}$ which may be
different.

 With such an ansatz for $f$, and noting that
\begin{equation}
\Delta E= Nk_B(T_z-T_\rho),
\end{equation}
eq. \ref{eq.evol} gives a differential equation for
$T_z-T_\rho$. For small initial deviations from equilibrium, we
can linearize the equation in $T_z-T_\rho$. With this
approximation, the evolution is exponential with a $1/e$ time
constant
\begin{equation}
T_{\rm therm}=
\frac{15\sqrt{\pi}}{2}\frac{1}
{\bar{n}\frac{8\pi\hbar^2}{\left ( \frac{m}{2}\right)^2\frac{k_BT_0}{m}}
\sqrt{\frac{k_BT_0}{m}}
}
\end{equation}
where $T_0$ is the equilibrium temperature and $\bar{n}$ is the
mean  density.

 We thus find that
\begin{equation}
\frac{1}{\bar{n}v_{rms}T_{\rm therm}}=\frac{2}{15\sqrt{\pi}}\sigma(v_{rms})
=\frac{64\sqrt{\pi}}{15}\frac{\hbar^2}{mk_B T_0},
\label{eq.3Dcoll}
\end{equation}
where $\sigma(v_{rms})$ is the cross section for a relative velocity
$v_{rms}=\sqrt{k_B T_0/m}$. 

As expected, the quantity
$\frac{1}{\bar{n}v_{rms}T_{\rm therm}}$ is inversely proportional to
the temperature. We have thus obtained the numerical factors entering
in  eq. \ref{eq.3Dcoll}, which is plotted in
fig.\ref{fig.collisionsExpetTheo}.

\subsection{Inhibition of thermalization when $k_{\rm B}T\ll \hbar \omega_{\rm osc}$}
\label{sec.inhib}

 We now give a simple physical interpretation of the exponential
decrease of the thermalization rate between the horizontal and
vertical degrees of freedom previously found in \cite{Petr01}.
We show that this behavior is due to the quantization of the energy
levels in the vertical direction and to parity conservation during the
collision process.

The collision rate is
\begin{equation}
\Gamma_{\rm coll}\simeq nv\sigma
\end{equation}
 where $n$ is the three dimensional density, $v$ is the typical
velocity of the atoms and $\sigma$ is the typical cross section.  For
$k_BT\ll \hbar\omega_{\rm osc}$, the horizontal velocity of the atoms is
much smaller than the typical vertical velocity
$\sqrt{\hbar\omega_{\rm osc}/(2m)}$ and the 3D density is
$n=n_{2D}\sqrt{2m/(\omega_{\rm osc} \hbar)}$, where $n_{2D}$ is the 2D
density.  Thus the collisional rate satisfies
\begin{equation}
\Gamma_{\rm coll}\simeq \frac{\hbar n_{2D}}{m}.
\end{equation}
$ \Gamma_{\rm coll}^{-1}$ is  an estimate of the thermalization time
between the two horizontal degrees of freedom, in agreement with the
rigorous calculation of ref.\cite{Petr01} as long as the
temperature is not extremely small \footnote{A logarithmic decrease of
the collision rate with the inverse of the temperature is expected
when $k_{\rm B}T\ll 0.05\hbar \omega_{\rm osc}$\cite{Petr00}.}.

\begin{figure}[ht]
\centerline{\includegraphics{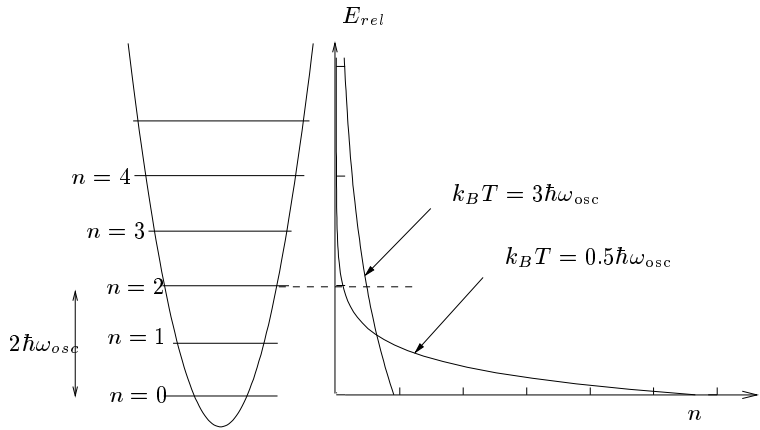}}
\caption{\it Probability distribution of the relative kinetic
energy of the horizontal motion of a pair of atoms for a
temperature of $3\hbar\omega_{\rm osc}/k_{\rm B}$ ( initial
conditions in our experiments) and a temperature of $0.5
\hbar\omega_{\rm osc}/k_{\rm B}$ ( end of 1D sideband cooling.}
\label{fig.distrienergie}
\end{figure}

On the other hand, the thermalization between the horizontal and the
vertical degrees of freedom actually takes  a much longer time.
Indeed, the energy levels are quantized in the vertical direction. In 
order to 
transfer energy to the vertical motion during a collision, the
horizontal kinetic energy of the atoms has to be at least
$\hbar\omega_{\rm osc}$ before the collision. In fact, because of parity
conservation, only the vibrational state $\ket{n=2}$ of the relative
motion can be populated after the collision\cite{Petr01}.  Thus,
the horizontal energy of the pair of colliding atoms has to be at
least $2\hbar\omega_{\rm osc}$.
 Figure \ref{fig.distrienergie} presents the probability distribution
$n(E)$ of the kinetic energy of the relative horizontal  motion of a
pair of  atoms, for a temperature of $3\hbar\omega_{\rm osc}$ and a
temperature of $0.5\hbar\omega_{\rm osc}$ close to the value achieved
in our experiments.  Because the density of states is constant in 2
dimensions, at a temperature $T$, $n(E)=1/(k_B T)exp(-E/(k_B T))$.
Thus, the probability  that a colliding pair has an energy large
enough to populate the excited states of the vertical motion is
\begin{equation}
p=e^{-\frac{2\hbar\omega_{\rm osc}}{k_B T}}.
\end{equation}

Among the collisions transferring energy to the vertical motion, we
can neglect the exponentially small amount that will transfer an
energy larger than $2\hbar\omega_{\rm osc}$. Thus, the rate of change of the
vertical energy is
\begin{equation}
\frac{dE_z}{dt}\simeq 2\hbar\omega_{\rm osc}
\frac{\hbar n_{2D}}{m}e^{-\frac{2\hbar\omega_{\rm osc}}{k_B T}}.
\label{eq.dezdt}
\end{equation}
 This exponential factor is responsible for the drastic inhibition
 of the cooling of the horizontal motion near $k_{\rm B}T \simeq \hbar \omega_{\rm osc}$ as already discussed in
 section \ref{sec.collhor}.

On the other hand, the increase of the vertical energy needed to
reach the thermal equilibrium is, neglecting the change of
horizontal temperature,
\begin{equation}
\Delta E_z=\hbar\omega_{\rm osc}e^{-\frac{\hbar\omega_{\rm osc}}{k_B T}}.
\end{equation}

Thus, the thermalization time $T_{\rm therm}$ is of the order of

\begin{equation}
\frac{1}{T_{\rm therm}}\simeq \frac{1}{\Delta E_z}\frac{dE_z}{dt}
\sim \frac{n_{2D}\hbar}{m}e^{-\frac{\hbar\omega_{\rm osc}}{k_B T}}
\end{equation}

 This formula is identical, within a numerical factor,
to the formula of ref.\cite{Petr01} using a quantum treatment of
collisions in 2D. It is interesting to note that both $dE_z/dt$
and $\Delta E_z$ are exponentially small but it is the difference
in the exponents that explains the exponential factor of $T_{\rm
therm}$.

The most efficient way to detect this exponential dependence of
$T_{\rm therm}$ would be to compare the thermalization time
between the two horizontal degrees of freedom ($\simeq \hbar
n_{2D}/m$)  to the thermalization time between them and the
vertical motion. However when the initial horizontal temperature
is $k_{\rm B}T\ll \hbar \omega_{\rm osc}$, even if the initial
vertical temperature is $\simeq 0$, the relaxation towards
equilibrium corresponds to an exponentially small change of the
horizontal temperature and an exponentially small population in
the excited vibrational states. The observation of these small
changes would require a very sensitive measurement of the
temperature or of the population of the vertical vibrational
states. Therefore the easiest choice of parameters to detect this
2D behavior is to operate near $k_{\rm B}T\simeq \hbar \omega_{\rm
osc}$.

\section{Summary}
In this paper we have investigated 1D sideband cooling of cesium atoms
in a far-detuned optical lattice. This cooling is particularly
efficient because, for each spontaneously emitted photon, an energy of
order $\hbar \omega_{\rm osc}$ is removed from the system. Atoms are
spin-polarized in F=3 and, at high density, elastic collisions enable
3D cooling of the sample through 1D sideband cooling. We have produced
atomic samples in which more than $80\%$ of the atoms are in the
vibrational ground state of motion in 1D and with a transverse
temperature of $0.7\, \hbar \omega_{\rm osc}/k_{\rm B}=2.8 \,\mu $K. This
realizes a quasi-2D cold gas.  Limitations of this cooling method have
been identified. First, light-induced atom losses have been
observed. Second, cooling of the weakly confined degrees of freedom by
collisions  looses its efficiency around $k_{\rm B} T\simeq \hbar
\omega_{\rm osc}$. We explained this effect by an exponential slowing
down of the energy transfer between the weakly and strongly confined
degrees of freedom. A dynamic method involving changes of the high
oscillation frequency via polarization rotation of one of the trapping
beams  has been proposed and implemented. We have shown that the zero
energy scattering resonance of cesium  was essentially unaffected in
the temperature range $4\,\mu $K--$20\,\mu $K. Within our experimental
accuracy, this is in agreement with theoretical calculations. Further
studies could concentrate on collisions affecting only the motion in
the weak confinement directions where dramatic modifications of
collisions occur at $k_{\rm B}T \leq 0.1 \hbar \omega_{\rm
osc}$\cite{Petr00,Petr01}. Promising systems for these studies are
Bose-Einstein condensates loaded in a far-detuned  trap associated
with the possibility of manipulating the scattering length via
Feshbach resonances.

%\section{Remerciements}
We gratefully thank Gora Shlyapnikov for help in theory and
interesting discussions.
Laboratoire Kastler Brossel is Unit{\'e} de recherche de l'Ecole Normale
Sup{\'e}rieure et de l'Universit{\'e} Pierre et Marie Curie, associ{\'e}e au CNRS. 
M. Morinaga thanks the University of Tokyo for its support. 

%\bibliography{bibartcoll}

\end{document}